\documentclass{article}
\usepackage{hiph-art}
\volnumber{19} \issuenumber{1} \edyear{2004}                             
\frompage{000} \topage{000}                                              
\recrevdate{1 January 2004}                                              
\def\rightmark{ }
\def\leftmark{ }
 
\title{From multifragmentation to supernovae and neutron stars} 
\authors{ 
{Ph.Chomaz$^1$, F.Gulminelli$^{2,a}$, 
 C.Ducoin$^{1,2}$, P.Napolitani$^1$ and K.Hasnaoui$^1$ %
\index{PhC} 
\index{FG} 
}\\[2.812mm]
{\normalsize
\hspace*{-8pt}$^2$ GANIL ( DSM - CEA / IN2P3 - CNRS),
B.P.5027, \\ F-14076 Caen C\'edex 5, France  
\hspace*{-8pt}$^1$ LPC Caen 
(IN2P3 - CNRS / EnsiCaen et Universit\'{e}), \\F-14050
Caen C\'{e}dex, France\\[0.2ex] 
}}
 
\abstract{The thermodynamics properties of globally neutral 
dense stellar matter are analyzed both in terms of mean field 
instabilities and structures beyond the mean field. 
The mean field response to finite wavelenght fluctuations is
calculated with the realistic Sly230a effective interaction.
A Monte Carlo simulation of a schematic lattice Hamiltonian 
shows the importance of calculations beyond 
the mean field to calculate the phase diagram of stellar matter.
The analogies and differences respect to the thermodynamics of 
nuclear matter and finite nuclei are stressed.}


 
\makeindex
\begin{document}
 
\maketitle

\section{Introduction}\label{intro}

Supernovae explosions, powered by the release 
of gravitational energy of a massive star which has exhausted its
nuclear fuel, can lead to the formation of a most interesting dense
stellar object: a neutron star\cite{yakovlev}. 
Due to the lack of observational data, 
the composition and structure of a neutron star is still highly 
hypothetical\cite{glendenning}. In the outer part of the star,
the stellar crust extending over about one kilometer, the matter 
density is expected to be comparable to normal nuclear matter density,
and the star can be modelized as essentially composed of neutrons,
protons, electrons and neutrinos in thermal and chemical equilibrium.
In a few minutes after its birth, the proto-neutron star formed at 
a temperature of $\approx 10^{11}K$ becomes transparent 
to neutrinos and cools via neutrino emission to temperatures which 
are small on the nuclear scale. 
The cooling process occurs via heat conduction and convection through
the envelope to the surface on a time scale too short for the system
to be in global thermal equilibrium, however local thermal equilibrium
should be well verified during the whole evolution of the 
proto-neutron star, and beta-equilibrium is also often
assumed\cite{lamb}. 
Crust matter is therefore very similar to nuclear 
matter, which is known 
to exhibit a complex phase
diagram including first and second order phase
transitions\cite{bertsch,serot}
.
 
The analogy between stellar matter and nuclear matter hides however
an important difference. 
If nuclear matter is by definition neutral,
only global charge neutrality is required by thermodynamic stability 
for stellar matter.
An important consequence of the charge neutrality constraint
is that the canonical free energy density $f$ is defined only for 
$\rho _{c} = \rho_{p}-\rho_{e}=0$. Hence $f\left( T,\rho_{n},\rho_{p},
\rho_{e} \right) = f\left( T,\rho_{n},\rho \right) $ 
with $\rho =(\rho_{p}+\rho_{e})/2$ 
and the chemical potential $\mu_c$ associated to $\rho_c$ 
can not be defined
since the free energy is not differentiable in the total-charge 
direction\cite{glendenning}. 
This constraint	affects the thermodynamics directly since it 
changes the number of degrees of freedom of the thermodynamic 
potentials\cite{paolo1}.
These considerations 
are expecially relevant when phase transitions are concerned.
For exemple, since $\mu_c$ 
is not a thermodynamic variable, 
the coexistence condition between two phases $A$ and $B$ 
$\mu^{A}=\mu^{B}$ does not imply that each of the 
chemical potentials $\mu_{e}$ and $\mu_{p}$ are identical in the 
two phases
. The difference in chemical potentials of charged particles
is counterbalanced by the Coulomb force:
as some electrons move from one phase to the other 
driven by the chemical-potential difference, the Coulomb force 
reacts forbidding a macroscopic charge to appear.

\section{Mean Field approximation}\label{meanfield}  
  
To illustrate the consequence of charge neutrality 
for the phase diagram of stellar matter, we present 
mean field calculations\cite{ducoin} using the Sly230a effective interaction,
which has been optimized to describe exotic nuclei and 
pure neutron matter\cite{chabanat}. This force has been already 
applied to neutron stars crust in ref.\cite{haensel}.

\subsection{Homogeneous matter}\label{homogeneous}

\begin{figure}[htb]
                 \insertplot{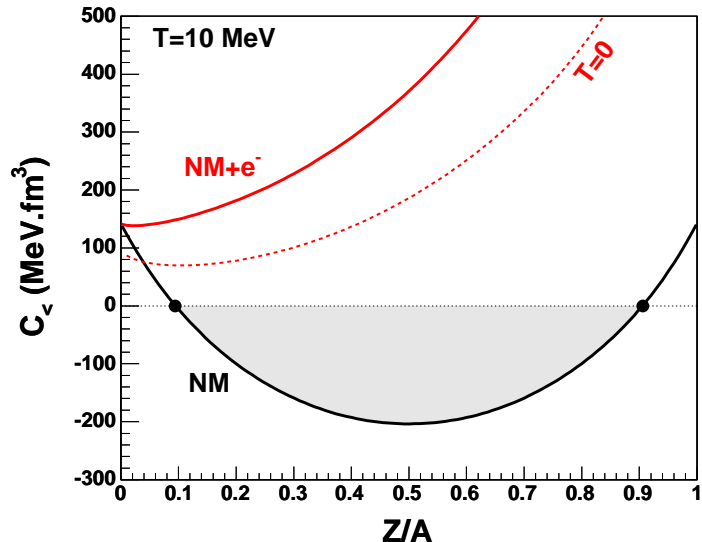}
\vspace*{-1.cm}
\caption[]{Minimum curvature of the mean field 
free energy matrix as a function of
isospin asymmetry for normal nuclear matter(NM) and for stellar matter
(NMe) at a temperature $T=10 MeV$ calculated 
with the Sly230a effective interaction. Dashed line:
stellar matter at zero temperature. }
\label{fig1}
\end{figure}

The matter instability to density fluctuations can be spotted 
looking at the curvature matrix\cite{pethick} 
\begin{equation}
	C =
	\left( 
	\begin{array}{ll}
	  \partial \mu_{n} / \partial \rho_{n} 
	& \quad\quad
	  \partial \mu_{n} / \partial \rho_{p} 
	\\ 
	  \partial \mu_{p} / \partial \rho_{n} 
	& 
	  \partial \mu_{p} / \partial \rho_{p} 
	+ \partial \mu_{e} / \partial \rho_{e}
	\end{array}
	\right) 
	\,\, ,\label{cmatrix}
\end{equation}
where the free energy density is
$f(T,\rho_{n},\rho ) = f_{N}(T,\rho _{n},\rho_p=\rho)
+f_{e}(T,\rho_e=\rho )$,
and we have introduced the chemical potentials $\mu_{n} =$  
$\partial f_{N}/\partial \rho_{n}$ , $\mu_{p} =$ $\partial f_{N}/ 
\partial \rho_{p}$ and $\mu_{e} =$ 
$\partial f_{e}/\partial \rho_{e}$. 
The additional term $\chi_{e}^{-1} =
\partial \mu_{e} / \partial \rho_{e}$ in the matrix 
modifies the stability conditions with respect to to the nuclear 
matter part, i.e. to the curvature of $f_{N}$. 
Due to the small electron mass, i.e. high Fermi energy, the
electron fluid is highly incompressible leading to 
a quenching of the instability: the instability conditions  
$\mathrm{Det}\,C\leq 0$ or $\mathrm{tr}\,C\leq 0$ are more 
difficult to fulfill.  
The result for the Sly230a interaction is shown in Figure \ref{fig1}:
the lowest eigenvalue of the curvature matrix is always positive
independent of the temperature and proton fraction, meaning 
that the spinodal zone is suppressed in stellar matter\cite{ducoin}.

\subsection{Finite wavelenght fluctuations}\label{finite}

\begin{figure}[htb]
                 \insertplot{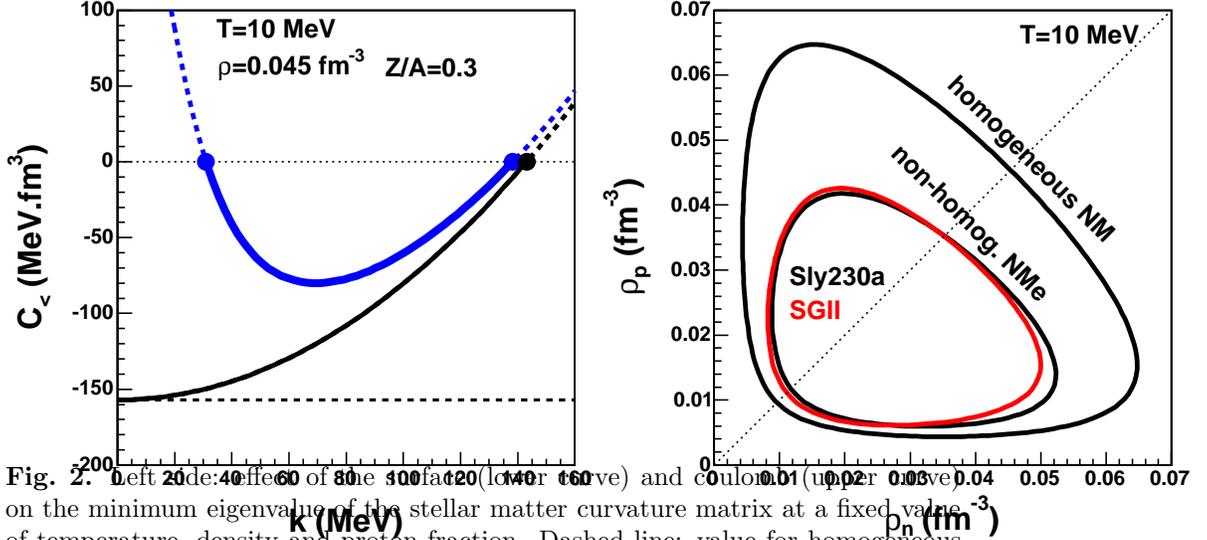}
\vspace*{-2.1cm}
\caption[]{Left side: effect of the surface (lower curve) and coulomb 
(upper curve) on the minimum eigenvalue of the stellar matter
curvature matrix at a fixed value of temperature, 
density and proton fraction. Dashed line: value for 
homogeneous nuclear matter.
Right side: instability region of homogeneous nuclear matter and 
inhomogeneous matter with electrons with two different effective
interactions.
 }
\label{fig2}
\end{figure}
 
If we expect stellar matter to be stable at all temperatures 
respect to global density fluctuations, it is well known that  
the matter ground state in the inner crust should correspond
to clusterized solid configurations composed of finite nuclei 
on a Wigner lattice\cite{negele}. To access these inhomogeneous
configurations, stellar matter may be instable respect to finite
wavelenght density fluctuations, $\delta \rho_q = A_q e^{i\vec k\cdot
\vec r} + A_q^* e^{-i\vec k\cdot\vec r}$, with $q=n,p,e$.
The mean field response to such a fluctuation induces two extra terms
in the curvature matrix eq.(\ref{cmatrix})\cite{pethick}.
The gradient term of the Skyrme fonctional\cite{chabanat}
produces a term $\propto k^2$ coupling proton and neutron densities,
which tends to suppress high $k$ fluctuations;
the direct Coulomb interaction produces a term $\propto k^{-2}$ 
coupling proton and electron densities,
which tends to suppress low $k$ fluctuations. The resulting 
effect 
is the appearence of a finite $k$ 
interval where one eigenvalue turns negative, i.e. matter becomes
unstable. An exemple is shown in Figure \ref{fig2}\cite{ducoin}.
We can see 
that an important instability region exists even at high temperatures 
and almost independent of the effective interaction employed, 
suggesting that clusterized configurations may be important also
in the proto-neutron star evolution.  

\subsection{The constraint of $\beta$-equilibrium}\label{beta}

\begin{figure}[htb]
                 \insertplot{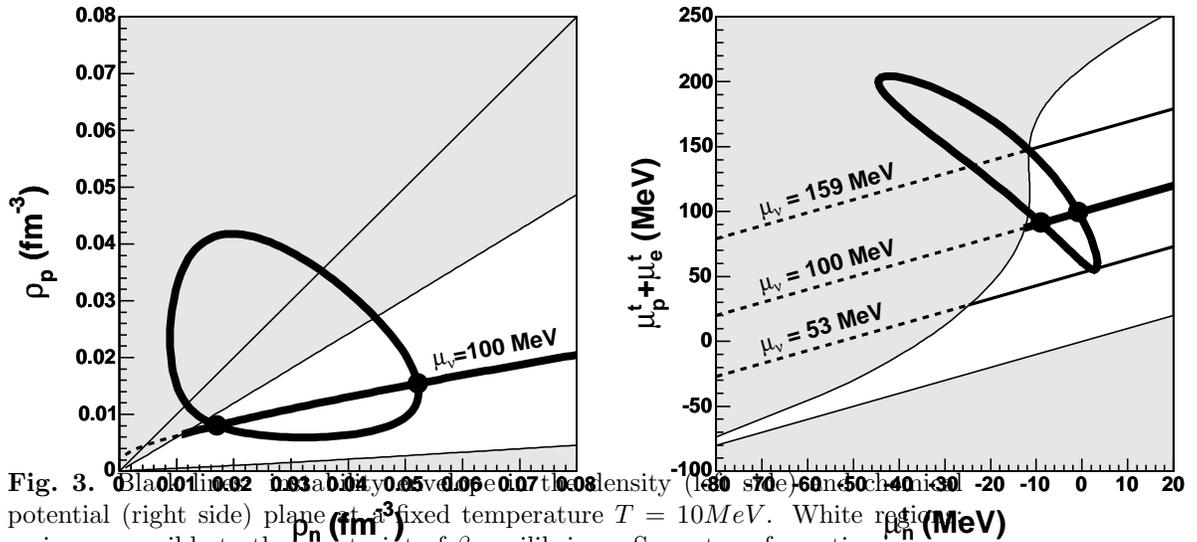}
\vspace*{-2.1cm}
\caption[]{Black lines: instability envelope in the density 
(left side) and chemical potential (right side) plane
at a fixed temperature $T=10 MeV$. 
White regions: regions accessible to the constraint of
$\beta$-equilibrium. Some transformations corresponding to 
different neutrino chemical potentials are also given.
 }
\label{fig3}
\end{figure}

If the cooling time scale is sufficiently slow, the proto-neutron 
star will also be subject to the constraint of $\beta$-equilibrium,
$\mu_p^{tot}+\mu_e^{tot}=\mu_n^{tot}+\mu_{\nu}$, 
with $\mu_q^{tot}=\mu_q+m_q$\cite{footnote1}. 
This is certainly 
the case in the final step of the star evolution, when the temperature
is low enough for the matter to be completely transparent to neutrino
emission, $\mu_\nu=0$, which becomes then the most effective cooling
process of the star\cite{yakovlev}.
Chemical equilibrium is generally assumed in the thermodynamic studies of
stellar matter\cite{pethick,haensel} and is often assimilated to
charge neutrality. However we would like to stress that 
chemical equilibrium, as well as any other other transformation
like constant proton fraction or constant temperature, 
is just a restriction of the accessible states and does not affect 
the thermodynamic properties, which are state functions\cite{footnote2}.

The chemical equilibrium constraint defines 
an accessible region in the phase diagram
shown for a given temperature in Figure \ref{fig3}\cite{ducoin}.
The lower limit is given by the equation $\mu_{\nu}=0$
while lepton number conservation defines an upper limit. 
We can see in Figure \ref{fig3} that, depending on $\mu_\nu$, i.e. on the 
neutrino opacity, the instability region can be crossed under the constraint
of chemical equilibrium.

\section{Coulomb frustration beyond the mean field}\label{frustration}
 
In the previous section we have seen that hot stellar matter 
can show an instability respect to finite wavelenght 
density fluctuations, eventually leading to the formation 
of clusterized configurations. This instability is due to the 
interplay between the surface and coulomb interaction terms.
This effect is a specific application of the generic physical concept
of frustration.
Frustration occurs in condensed matter physics whenever 
interactions with opposite signs act on a comparable lenght scales;
applications range from magnetic systems to liquid crystals, from spin
glasses to protein folding\cite{Tarjus}.
A well known application of frustration in nuclear physics is given
by the multifragmentation observed in violent ion collisions. 
Concerning compact stars, it is recognized since a long time
that in the clusterized state generated by frustration 
the structures may abandon spherical shapes and organize according 
to more complex topologies (pasta phases)
\cite{Ravenhall}. 
The possible survival of these structures at high temperature 
is discussed in the recent literature\cite{Watanabe,Horowitz}.

\subsection{A statistical treatment of frustration}\label{stat}

\begin{figure}[htb]
                 \insertplot{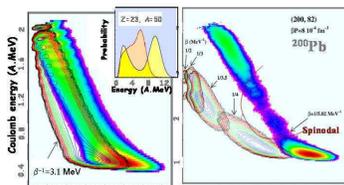}
\vspace*{-2.1cm}
\caption[]{ Event distribution of the SMM model in the Coulomb energy 
versus total energy plane for a finite nucleus of atomic number $Z=23$ 
(left part) and $Z=82$ (right part) in the multi-canonical ensemble.
Contour plots: uncharged system $\beta_C=0$; contour lines:
charged system $\beta_C=\beta_N$ with different values of $\beta_N$.
Full line: spinodal curve. Insert: projection over the total energy axis
for the uncharged (bimodal distribution) and the charged (monomodal
distribution) case.
 }
\label{fig4}
\end{figure}

A statistical tool to deal with frustrated systems 
is given by the multi-canonical ensemble\cite{raduta}.
The two energy components $E_N,E_C$ are treated as two independent 
observables associated to two Lagrange multipliers
$\beta_N,\beta_C$. A generalized grand potential is defined by
 
\begin{equation}
Z_{\beta_N,\beta_C,\vec\alpha}=\int d\vec N Z^c_{\beta_N,\beta_C}(\vec N)
exp \left (-\vec{\alpha}\cdot \vec{N} \right ) \, , \label{macro}
\end{equation}

where $e^{\alpha_q}$ is the fugacity of particle type $q$ 
and the multi-canonical partition sum reads

\begin{equation}
Z^c_{\beta_N,\beta_C}(\vec N)=\int dE_N dE_C W(E_N,E_C,\vec N)
exp \left ( -\beta_N E_N -\beta_C E_C \right )  \, . \label{cano}
\end{equation}
 
If $E_C$ represents the Coulomb energy and $E_N$ the nuclear term,
the choice $\beta_C=\beta_N$ gives the usual (grand)canonical 
thermodynamics for charged systems, $\beta_C=0$ leads to the uncharged
thermodynamics, while all intermediate values $0<\beta_C<\beta_N$
correspond to interpolating ensembles, or equivalently
to physical systems with an effective charge 
$(q_{eff}/q_0)^2=\beta_C/\beta_N$. The  
multi(grand)canonical ensemble allows to construct a single unified 
phase diagram for neutral and charged matter, and is therefore an ideal
statistical tool to make some connections between nuclear and 
stellar matter.  
 
\subsection{Application to multifragmentation}\label{raduti}

As an example let us consider the phenomenon of 
multifragmentation as modelized in the Statistical Multifragmentation 
Model\cite{raduta}. 
The event distribution in the multi-canonical ensemble eq.(\ref{cano})
is represented 
in Figure \ref{fig4} for a value $\beta_N$ corresponding to the transition
temperature for the uncharged case.
The 
first order phase transition gives rise to a two peaked
distribution in the (multi)canonical ensemble, representing the two
coexisting phases. The direction separating the two peaks can be taken
as a definition of the order parameter 
\cite{topology}.
From the left part of Fig.\ref{fig4} we can see that the effect of the 
Coulomb interaction is a rotation of the order 
parameter which becomes almost orthogonal to the total energy direction,
leading to a reduction of the latent heat and a shrinking 
of the coexistence zone. When the Coulomb effect becomes important
(right part), the spinodal region 
is not explored by the charged system 
and the phase transition becomes a cross-over.

\subsection{An Ising based model of stellar matter}\label{ising}

\begin{figure}[htb]
                 \insertplot{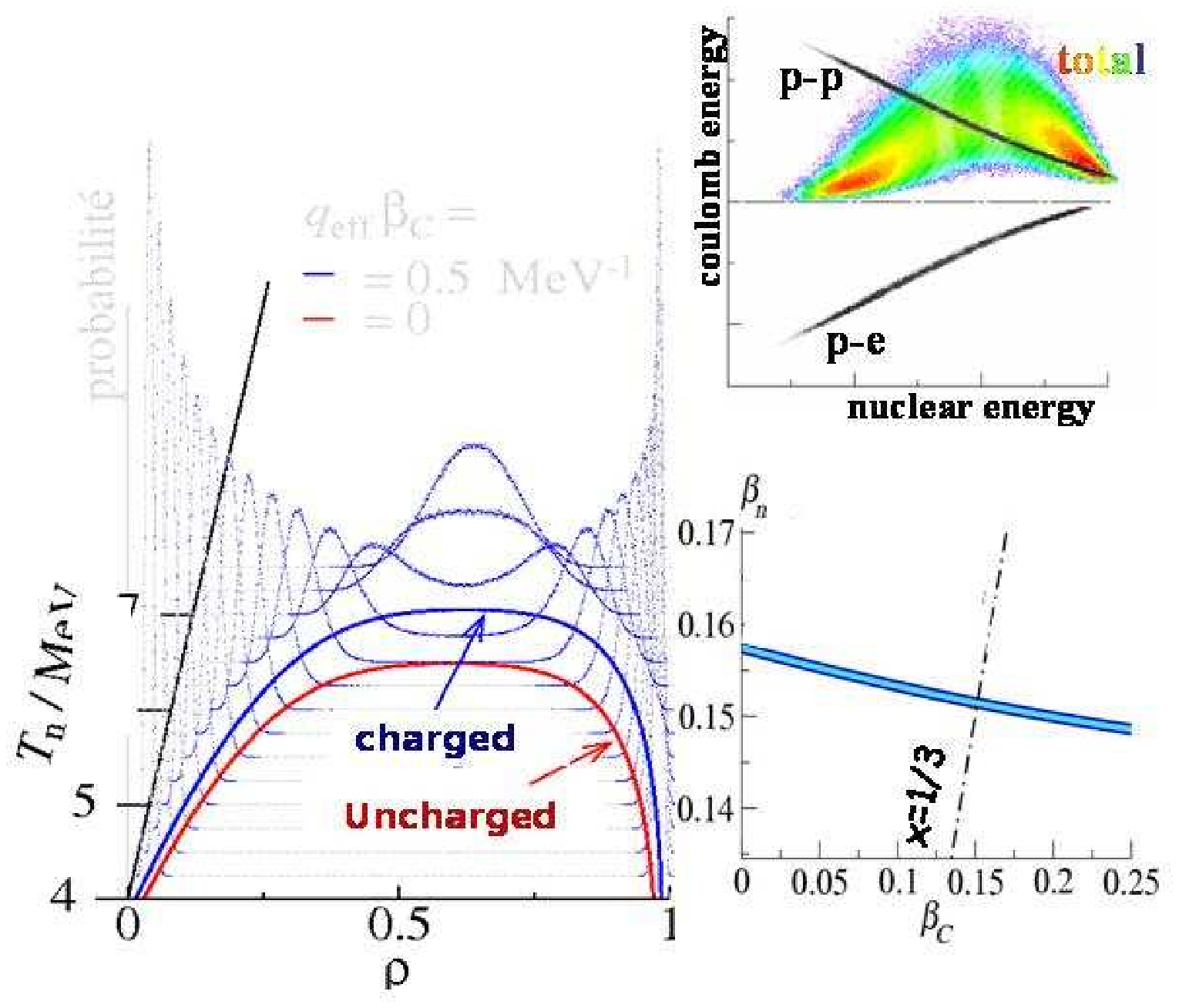}
\vspace*{-1.1cm}
\caption[]{ Upper right: typical low temperature event distribution
of the long ranged globally neutral Ising model. The proton-proton, 
proton-electron, and total distributions are given. 
Left: coexistence region for the uncharged and charged case obtained 
from the multi-grand-canonical particle number distributions.
Lower right: phase diagram. The constant proton fraction $x=1/3$ line
is also shown.
 }
\label{fig5}
\end{figure}

To get a qualitative information of the effect of frustration
on dense globally neutral stellar matter we have introduced a 
schematic but exactly solvable Ising model with long Coulomb-like 
and short nuclear-like range interactions\cite{paolo2}
 
\begin{equation}
H_N = -\frac{\epsilon}{2}\sum_{<ij>}n_i n_j \;\; ; \;\;  
H_C =   \frac{\chi}{2}\sum_{i\neq j} \frac{q_i q_j}{r_{ij}}
         =\frac{\chi}{2}\sum_{i\neq j} n_i n_j C_{ij} \, ,
\end{equation}
 
where each site of a three dimensional lattice of $N=L^3$ sites  
is characterized by an occupation number $n_i=0,1$ and an effective
charge $q_i=n_i-\sum_{j=1}^N n_j/N$. The effective charge represents the 
proton distribution screened by an uniform electron 
background\cite{Magierski}.
To accelerate thermodynamic convergence, the finite lattice is repeated
in all three directions of space a large number $N_R$ of times

\begin{equation}
H_C^{tot} = \frac{\chi}{2}\sum_{n=1}^{N_R}
\sum_{i,j}^N n_i n_j C_{IJ}
\end{equation}

with $\vec I=\vec i + \vec n L$, $\vec J=\vec j + \vec n L$.


\begin{figure}[htb]
                 \insertplot{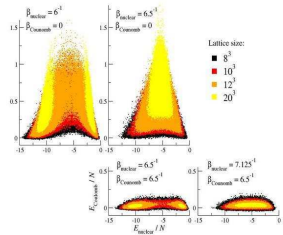}
\vspace*{-1.1cm}
\caption[]{Event distribution of the long ranged globally neutral Ising model
in the Coulomb energy versus nuclear energy plane for different values 
of the Lagrange multipliers and different lattice sizes.  
 }
\label{fig6}
\end{figure}

The resulting phase diagram is shown in Figure \ref{fig5}.
We can see that, at variance with the case of multifragmentation of 
finite charged nuclei, the effect of the Coulomb interaction is 
an expansion of the coexistence zone. This  result
can be easily understood inspecting the event distributions in the 
multicanonical ensemble in Fig.\ref{fig5}\cite{paolo2}. We can see that 
the proton-proton and proton-electron energy distributions are opposite
in sign and similar in shape; in fact they would be exactly identical
in the mean field approximation, meaning that the only effect of the 
charge at the mean field level is the addition of the electron free energy
as we have discussed in section \ref{meanfield}. The sum of the two
contributions in this exactly solved model 
gives a two peaked distribution indicating coexistence between
a high nuclear energy liquid and a low nuclear energy gas.
The electron screening effect minimizes the Coulomb energy of the dense phase,
making it accessible even at high temperature at variance with the finite
nucleus case. This result implies that pasta phases may be relevant in
a wide range of thermodynamic conditions
.

Another important effect of the Coulomb 
interaction in stellar matter is to suppress the critical point 
of neutral nuclear matter and the related phenomenon of 
critical opalescence.
The Coulomb energy density can be expressed as 
\begin{equation}
	\frac{\langle\hat{V}_{c}^{\prime }\rangle}{V}=\frac{\alpha }{2V}\int 
	\frac{\sigma_c (\mathbf{r},\mathbf{r}^{\prime })}
	{\left| \mathbf{r}-\mathbf{r}^{\prime }\right|} 
	\mathrm{d} \mathbf{r}\mathrm{d} \mathbf{r}^{\prime }=2\pi \alpha \int 
	\sigma_{c}(r) r \mathrm{d} r 
	\,\, ,
	\label{crit}
\end{equation}
where $\sigma (\mathbf{r},\mathbf{r}^{\prime }) = 
\langle\delta \rho_{c}(\mathbf{r})\delta \rho_{c}(\mathbf{r}^{\prime })\rangle$ 
is the charge density fluctuation. 
At the critical point 
$\sigma _{c}(r)$ is expected to 
scale as $\sigma _{c}(r)\propto r^{-1-\eta } $, 
where 
$\eta$ is a critical
exponent which turns out to be close to zero in most physical 
systems. 
A critical point would then correspond to a divergent 
Coulomb energy. 
As a consequence, the phase transition in stellar matter ends	
at a first-order point.
This can be clearly seen in our numerical simulation as shown in 
Figure \ref{fig6}\cite{paolo2}. The critical point of the uncharged system
is characterized by a diverging coulomb energy (upper right), while
for the charged system  at the end point of the coexistence zone 
(lower right) the fluctuation does not increase with the 
lattice size.

Such an effect has been already observed in Ising models  
with long-range frustrating interactions, where the coexistence  
region was seen to end at a first-order point~\cite{Tarjus}.
Since the Fourier transform of the correlation function 
gives the enhancement of the in medium scattering cross section 
respect to its free value, the important physical implication of this 
result is that we expect hot stellar matter to show a small opacity 
to neutrino scattering\cite{Horowitz}, which may  
have important consequences on the cooling dynamics \cite{Margueron}. 
 
\section{Conclusions}\label{concl}

In this contribution we have estabilished some connections 
between exotic phases in stellar matter and nuclear multifragmentation.
Both phenomena are triggered by an instability to finite wavelenght 
density fluctuations and can be understood in terms of frustration 
between the nuclear and the Coulomb interaction.

If the first order phase transition of normal nuclear matter 
is suppressed in stellar matter due to the high electron 
incompressibility, a (second order\cite{paolo1}) transition 
is however expected between single-phase and coexistence 
configurations.   
Large correlated structures are therefore predicted,
even if at the ending point of the coexistence region the 
correlation lenght is not expected to diverge.

Since the Coulomb energy is minimized in 
homogeneous partitions corresponding to pure phases, and it is
maximal in clusterized partitions, a higher 
temperature is needed to reach the mixed partitions 
with respect to the uncharged system. 
Therefore, the mixed-phase phenomenology may be relevant 
for the proto-neutron-star structure 
in a wider temperature range than usually expected~\cite{glendenning}.
Such an expansion of the coexistence region is at
variance with the mean-field expectations 
. This contrast stresses the importance of 
realistic calculations beyond
the mean-field level~\cite{Watanabe,Horowitz}. 
  
\section*{Notes}  
\begin{notes}
\item[a]
Member of the Institut Universitaire de France\\ 
\end{notes}

\vfill\eject

\begin{thebibliography}{99}
\bibitem{footnote1} Since the density dependence of the neutron and
proton effective mass is poorely known, we have used bare masses in this
expression.
\bibitem{footnote2}Conversely, charge neutrality 
affects the thermodynamics 
directly since it changes the number of 
degrees of freedom of the thermodynamic potentials.
\end{thebibliography}
\end{document}